\documentclass[conference]{IEEEtran}

\usepackage{cite}
\usepackage{amsmath,amssymb,amsfonts}
\usepackage{algorithmic}
\usepackage{graphicx}
\usepackage{textcomp}
\usepackage{xcolor}
\usepackage{url}
\usepackage{booktabs}
\usepackage{siunitx}

\usepackage{siunitx}
\usepackage{hyperref}
\usepackage[nolist]{acronym}
\usepackage[flushleft]{threeparttable}
\usepackage{placeins}
\usepackage{svg}                                   
\usepackage[percent]{overpic}

\usepackage{mathtools}
\usepackage{multirow}
\usepackage{booktabs}
\usepackage{subcaption}

\newcommand{\RNum}[1]{\uppercase\expandafter{\romannumeral #1\relax}}   

\DeclareRobustCommand{\IEEEauthorrefmark}[1]{\smash{\textsuperscript{\footnotesize #1}}}

\DeclareSIUnit\db{dB}                           
\DeclareSIUnit\dbi{dBi}                         
\DeclareSIUnit\dbm{dBm}                         
\DeclareSIUnit\watthour{Wh}                     
\DeclareSIUnit\msInference{ms/inference}        

\def\BibTeX{{\rm B\kern-.05em{\sc i\kern-.025em b}\kern-.08em
    T\kern-.1667em\lower.7ex\hbox{E}\kern-.125emX}}

    \usepackage{fancyhdr}   

\fancypagestyle{firstpage}{     
  \fancyhf{}
  \chead{\textcolor{gray}{This article has been accepted for publication in the flagship conference of \\ the IEEE International Instrumentation and Measurement Technology Conference(I2MTC).}}
  \fancyfoot[C]{\small{\textcolor{gray}{~\copyright~ 2025 IEEE.  Personal use of this material is permitted.  Permission from IEEE must be obtained for all other uses, in any current or future media, including reprinting/republishing this material for advertising or promotional purposes, creating new collective works, for resale or redistribution to servers or lists, or reuse of any copyrighted component of this work in other works.}}}

}

\begin{document}

\title{Smart Feeding Station: Non-Invasive, Automated IoT Monitoring of Goodman's Mouse Lemurs in a Semi-Natural Rainforest Habitat\\}

\author{
\IEEEauthorblockN{
Jonas Peter\IEEEauthorrefmark{1}, 
Victor Luder\IEEEauthorrefmark{1},
Leyla Rivero Davis\IEEEauthorrefmark{2}, 
Lukas Schulthess\IEEEauthorrefmark{1}, 
Michele Magno\IEEEauthorrefmark{1},
}\\

\IEEEauthorblockA{\IEEEauthorrefmark{1} Dept. of Information Technology and Electrical Engineering, ETH Zurich, Switzerland}
\IEEEauthorblockA{\IEEEauthorrefmark{2} Zoo Zurich, Dept. for Research and Species Conservation, Switzerland}
}
\maketitle

\begin{abstract}
In recent years, zoological institutions have made significant strides to reimagine ex situ animal habitats, moving away from traditional single-species enclosures towards expansive multi-species environments, more closely resembling semi-natural ecosystems. This paradigm shift, driven by a commitment to animal welfare, encourages a broader range of natural behaviors through abiotic and biotic interactions. This laudable progression nonetheless introduces challenges for population monitoring, adapting daily animal care, and automating data collection for long-term research studies. This paper presents an IoT-enabled wireless smart feeding station tailored to Goodman's mouse lemurs (\textit{Microcebus lehilahytsara}). System design integrates a precise \ac{RFID} reader to identify the animals' implanted \ac{RFID} chip simultaneously recording body weight and visit duration. Leveraging sophisticated electronic controls, the station can selectively activate a trapping mechanism for individuals with specific tags when needed. Collected data or events like a successful capture are forwarded over the \ac{LoRaWAN} to a web server and provided to the animal caretakers. To validate functionality and reliability under harsh conditions of a tropical climate, the feeding station was tested in the semi-natural Masoala rainforest biome at Zoo Zurich over two months. The station detected an animal's \ac{RFID} chip when visiting the box with $\mathbf{98.68\mskip3mu}$\% reliability, a \ac{LoRaWAN} transmission reliability of $\mathbf{97.99\mskip3mu}$\%, and a deviation in weighing accuracy below $\mathbf{0.41\mskip3mu}$g. Beyond its immediate application, this system addresses the challenges of automated population  monitoring advancing minimally intrusive animal care  and research on species behavior and ecology.

\end{abstract}

\vspace{10pt}

\section{Introduction}
\thispagestyle{firstpage}

 Modern zoos are increasingly investing in animal welfare and shifting to spacious, semi-naturalistic immersive exhibits where multiple species co-exist in environments that resemble their natural habitats\cite{vision_zürich_zoo}~\cite{zoo_welfare}. This shift necessitates novel animal monitoring technologies~\cite{j:liptovszky_digital_animal_welfare} particularly for cryptic species, in complex habitats where animal tracking and visibility is a challenge. A routine population census allows zookeepers and veterinarians to address  individual  animal health needs and collect baseline data for long-term research furthering understanding of understudied species biology and behavior.

However, a zoo's resources to invest in zoo staff executing such work are limited. Data collection, processing, and equipment maintenance are often labor-intensive and require time and effort from staff. These circumstances highlight the need for easy-to-use, low-maintenance automated systems that integrate into existing zookeeper workflows~\cite{berckmans_2017}. 

With the rise of \ac{IoT}, such systems may consist of multiple sensors embedded into the zoo's environment, streamlining automated data collection and ultimately increasing the level of digitization in zoos ~\cite{j:liptovszky_digital_animal_welfare}. The design can be of various forms. They may be interactive, aiming to engage animals on multiple tasks while data is collected ~\cite{brooks_2022}. Other goals are tracking and localization methods through sound localization of specific individual animals ~\cite{schneider_2021}. Lastly, smart feeding stations may capture regular visits by  animals' and can be a valuable juncture for data collection ~\cite{Harahap_2022}. A key criterion for such systems is to be noninvasive to minimize any impact on animals~\cite{wolfensohn_2018}.
However, expanding the monitoring coverage across the zoo presents challenges since species-specific idiosyncrasies may often require species-tailored monitoring solutions ~\cite{binding_2020}.

One case study species  is the Goodman's mouse lemur (\textit{Microcebus lehilahytsara}), a species recognized as distinct only in 2005 by Kappeler et al.~\cite{dammhahn_2005} and brought into human care ex situ in 2007. Native to Madagascar's eastern rainforests and the Central Highlands~\cite{tiley_2021}, these lemurs are of interest due to the novelty of their discovery. At Zurich Zoo, the Masoala rainforest, which resembles Madagascar's environment with an expansive 11000 square meter space~\cite{masoala_2024}, is currently home to 126 mouse lemurs (ZIMS, 1. March 2025). This naturalistic setting allows the lemurs to move freely. However, it raises questions about their social dynamics and well-being in such a spacious, open environment. This setup is an ideal application area for an innovative, automated monitoring system since it allows gathering insightful data in an environment resembling the animals' real habitat.  

This paper presents the design and implementation of an IoT-enabled feeding station to monitor the behavior of Goodman's mouse lemurs. The system is designed to record data on their visits' frequency, timing, and location. It also collects well-being indicators by measuring each lemur's weight and eating habits. Additionally, it includes an automated mechanism to selectively capture individual lemurs, targeting to support the annual health checks conducted by zookeepers. In combination with the \ac{LoRaWAN} network, provided by EWZ (Elektrizitätswerk der Stadt Zürich, provider of a city-wide \ac{LoRaWAN} network), the data are collected and stored on a centralized server. A custom web interface makes the data accessible and reduces the needed data collection effort, ultimately accelerating and streamlining the process. 

The proposed system accounts for several challenges. First, harsh environmental conditions, including daily artificial rainfalls, require robust electronic components and adequate protection. Second, the design prioritizes low maintenance, with features to ensure easy cleaning and accessibility for zoo staff. Finally, it meets the researchers' requirements of providing accurate, reliable, high-quality data with minimal additional effort.

Within this context, our solution includes the following strengths and contributions:

\begin{itemize}
\item The design and implementation of an end-to-end LoRa-based sensing system developed to monitor Goodman's mouse lemurs in an ex situ habitat allow zookeepers and researchers to collect more detailed and seamless insights into the animals' behavior and health. The high level of automation streamlines the data collection and minimizes additional maintenance efforts. Moreover, the system can be integrated into any zoological environment with minimal setup requirements. The system is designed to read the \ac{RFID} tags on animals, identify untagged individuals, and record their weight. In addition, it assists zoo staff in capturing animals by automating this process, allowing selective trapping based on tag number.

\item Accurate evaluation of all functionalities tested in an actual application under harsh external conditions given by the tropical environment of the Masoala rainforest. This test proved the setup to be robust and enduring, rendering it suitable for such a use case. 

\item The fully automated, end-to-end system streamlines data collection and provides server-side processing that makes the collected data accessible. This system illustrates the potential impact on zoos, as it integrates modern animal care concepts, such as large, open habitats, with non-invasive data-based research on animal behavior and their well-being.
\end{itemize}

\section{Related Work}
Various approaches have been explored in animal monitoring to collect data in zoological environments. For example, Asher et al. ~\cite{asher_2018} studied the well-being of elephants in zoos based on the collection of behavioral indicators. However, their approach required manual assessments by zookeepers observing the elephants, significantly increasing the workload. 
Camera-based systems have been tested in more traditional enclosures, where animals move within defined spaces, and their hiding opportunities are limited. For instance, Zuerl et al. ~\cite{zuerl_2024} developed an automated monitoring system based on cameras to reduce the labor-intensive task of observing polar bears and their activities. Similarly, Shannon et al. ~\cite{shannon_2024} used cameras to optimize evidence-based care for aquatic turtles. 
Nevertheless, a sensor fusion approach, combining camera data with \ac{RFID} tag scanning, was implemented by Karlsson and colleagues ~\cite{karlsson_2010} to monitor animal locations and activities across an entire zoo. While most current monitoring systems primarily focus on localization, little attention has been given to collecting other parameters, such as weight. An example of such a non-invasive, however complex, system for animal weight measurement is presented by Larios ~\cite{Larios_2013}. 
On the other hand, limited research has been conducted on Goodman's mouse lemurs in zoological environments. Meyer et al. ~\cite{meyer_2013} studied the social behavior of this species by tracking ten individuals using radio collars. A similar study, based on the same technology, was conducted in Madagascar by Andriambeloson and his colleagues ~\cite{Andriambeloson_2021}. They tracked two lemurs in their natural habitat for several days. However, overall research on this species remains sparse, as well as the deployment of automated, non-invasive animal tracking setups aiming to support zoos in their research and operation. 
Building on these previous approaches, our work introduces a novel IoT-enabled smart feeding station explicitly designed for Goodman's mouse lemurs. Unlike existing systems, our setup identifies individual animals and monitors their activity and seamlessly integrates weight measurement and selective interaction capabilities in a single, automated platform. Accurate measurements collected in real-world field conditions are presented and thoroughly discussed, addressing the challenges of precision in ecological monitoring. 

\section{System Overview}

The design of the feeding station must address two primary aspects. First, the physical setup has to be composed of durable material capable of withstanding the harsh environmental conditions of a rainforest. This includes the protection of the electronics against periodic rainfall, humidity levels reaching up to 100\%, corrosion, and potential infestation by ants. To meet these requirements, the hardware must ensure complete dust protection, water resistance, and moisture immersion. Nonetheless, the zookeepers' working routine must also be considered, as the system specifically aims to reduce their workload. Therefore, the design must prioritize easy cleanability and streamlined data access, making the collected information available for research without any additional steps. To give an overview, each sub-system of the designed hardware and software is described in the following sections.

\begin{figure}
    \centering
    \includegraphics[width=0.8\linewidth]{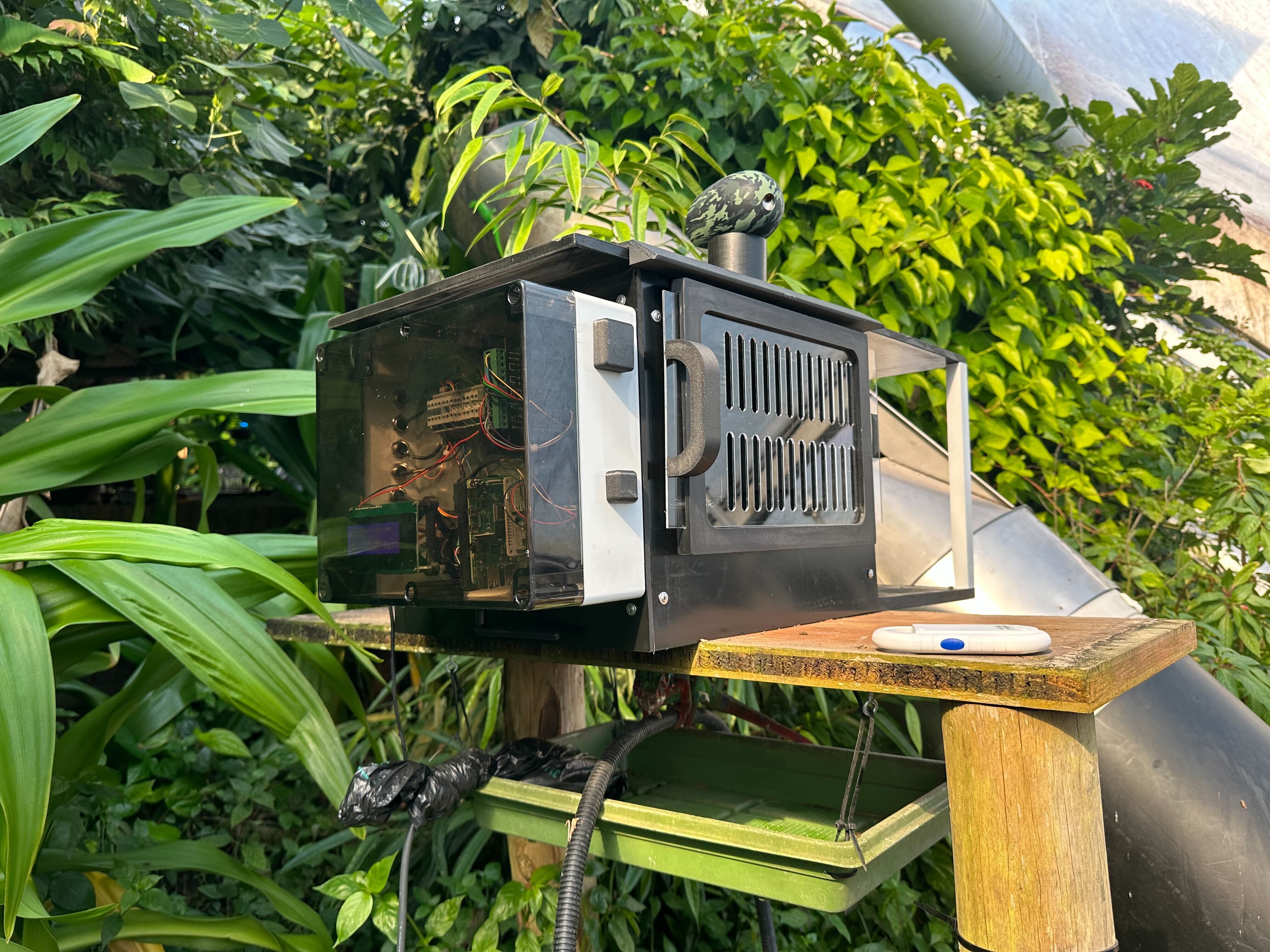}
    \caption{Final deployment of the smart feeding station at the zoo's Masoala rainforest.}
    \label{fig:dployed_box}
\end{figure}

\subsection{Hardware}
\begin{figure}
    \centering
    \includegraphics[width=0.9\linewidth]{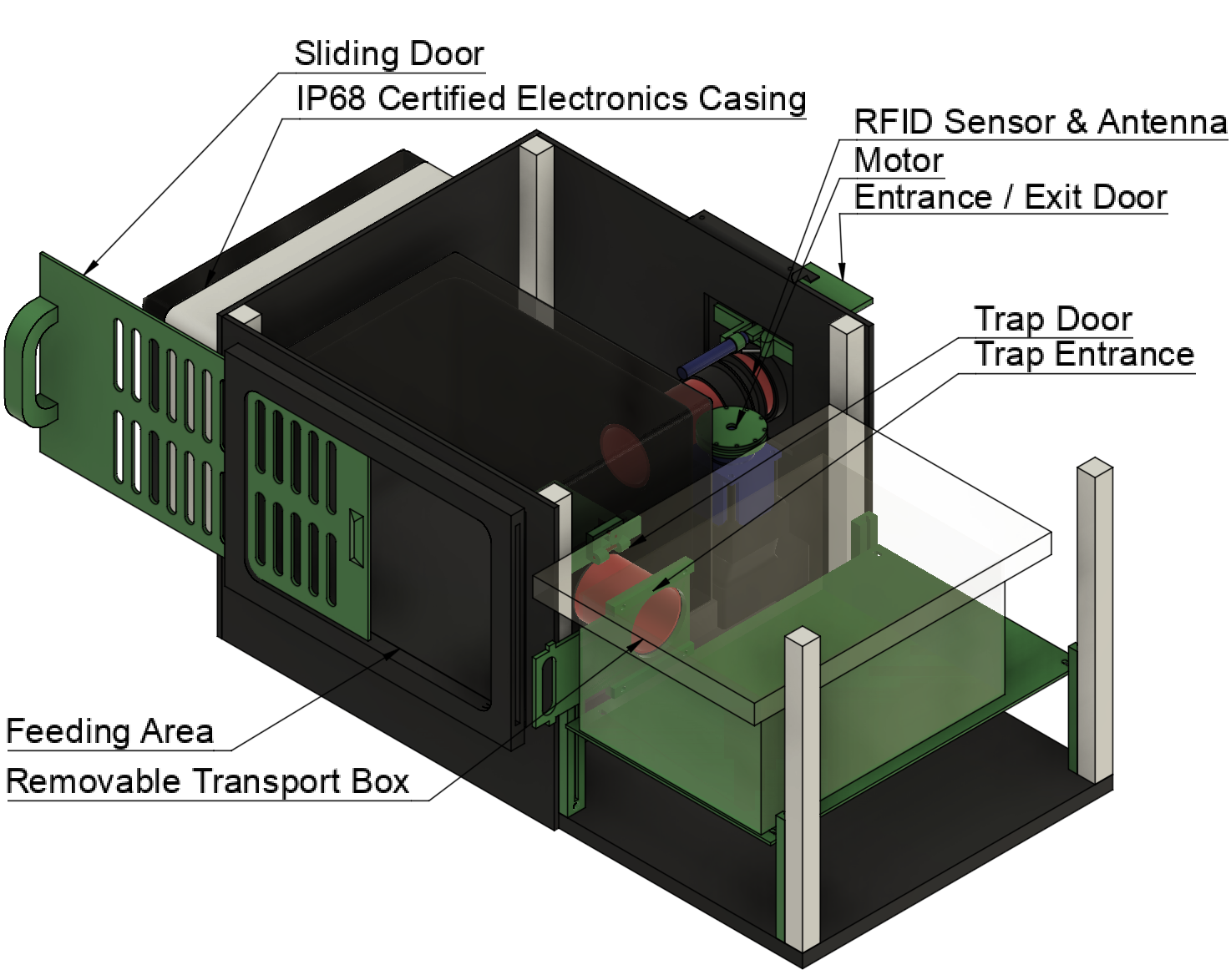}
    \caption{Overview of the final system with roof and side panel removed.}
    \label{fig:sys_overview}
\end{figure}
  
\subsubsection{Feeding Station Design}
As illustrated in Figure \ref{fig:sys_overview}, the feeding station consists of an IP68-certified electronics casing chosen for robust protection. It holds the Raspberry Pi computation module, which controls the components, including the motor controller, \ac{RFID} scanner, LoRa data transmission module, and sensors. This includes an \ac{ADC} to read the load cell and the humidity sensors to monitor environmental conditions inside and outside the casing. The connected modules are illustrated in Figure \ref{fig:software_overview}. Furthermore, the station incorporates a feeding area accessible through an entry tube equipped with an \ac{RFID} antenna. The feeding area is made of a single unit using 3D printing, which improves cleanability. It is mounted to a metal structure equipped with a load cell to measure the animals' weight.
Additionally, the feeding area incorporates a trap system designed to guide animals into the entry tube or a trap tube, leading to a transport box. This functionality is controlled by a stepper motor that actuates doors mounted on both tubes. The motor simultaneously closes the entry tube while opening the trap tube. This setup allows targeted trapping of animals while keeping the feeding area accessible for others.

\subsubsection{Weighing System}
The weighing system comprises the Sparkfun NAU7802 \ac{ADC} converter and the Zemic L6D \qty{6}{\kilo\gram} load cell. The \ac{ADC} features a resolution of 24 bits, and its gain is set to 128, while samples are collected with a frequency of \qty{20}{\hertz}. The basis of the weighing platform is a metal frame with a size of \qty{300}{\milli\meter} x \qty{300}{\milli\meter}. To improve the robustness and accuracy of the weight measurements, each part where an animal could be located, such as the feeding area and entrance tube, is placed onto the metal frame. This resolves the problem of an animal only being partially on the platform and improves the system's cleanability as gaps between weighing and non-weighing areas are reduced.

\subsubsection{RFID Scanner}
The \ac{RFID} scanner and antenna are located at the beginning of the entrance tube. An ID-3LA-ISO scanner with a custom antenna matching the required inductance is used for the scanning. This system is suited for the commonly used FDX-B animal identification protocol, which is based on the ISO 11784 \& 11785 standards. 

\subsubsection{LoRaWAN Integration}
For wireless data transmission and automated data collection, the feeding station is integrated into the LoRa network of EWZ. On the hardware side, an extension module featuring the Rak811v3 transceiver module is connected via UART to the Raspberry Pi.  

\subsubsection{Trapping Mechanism}
The two doors of the system are spring-loaded. This ensures that the force of the motor is not acting on the door but is defined by the spring. Preventing injuries in case an animal is pinched. The state of the doors is determined using an inductive proximity switch, which provides feedback for auto-calibration. Only one motor and proximity switch are incorporated through the mechanically coupled doors, which allows the system to remain lightweight and limits the number of components and cables exposed to high humidity.

\subsection{Software}
\begin{figure}
    \centering
    \includegraphics[width=1\linewidth]{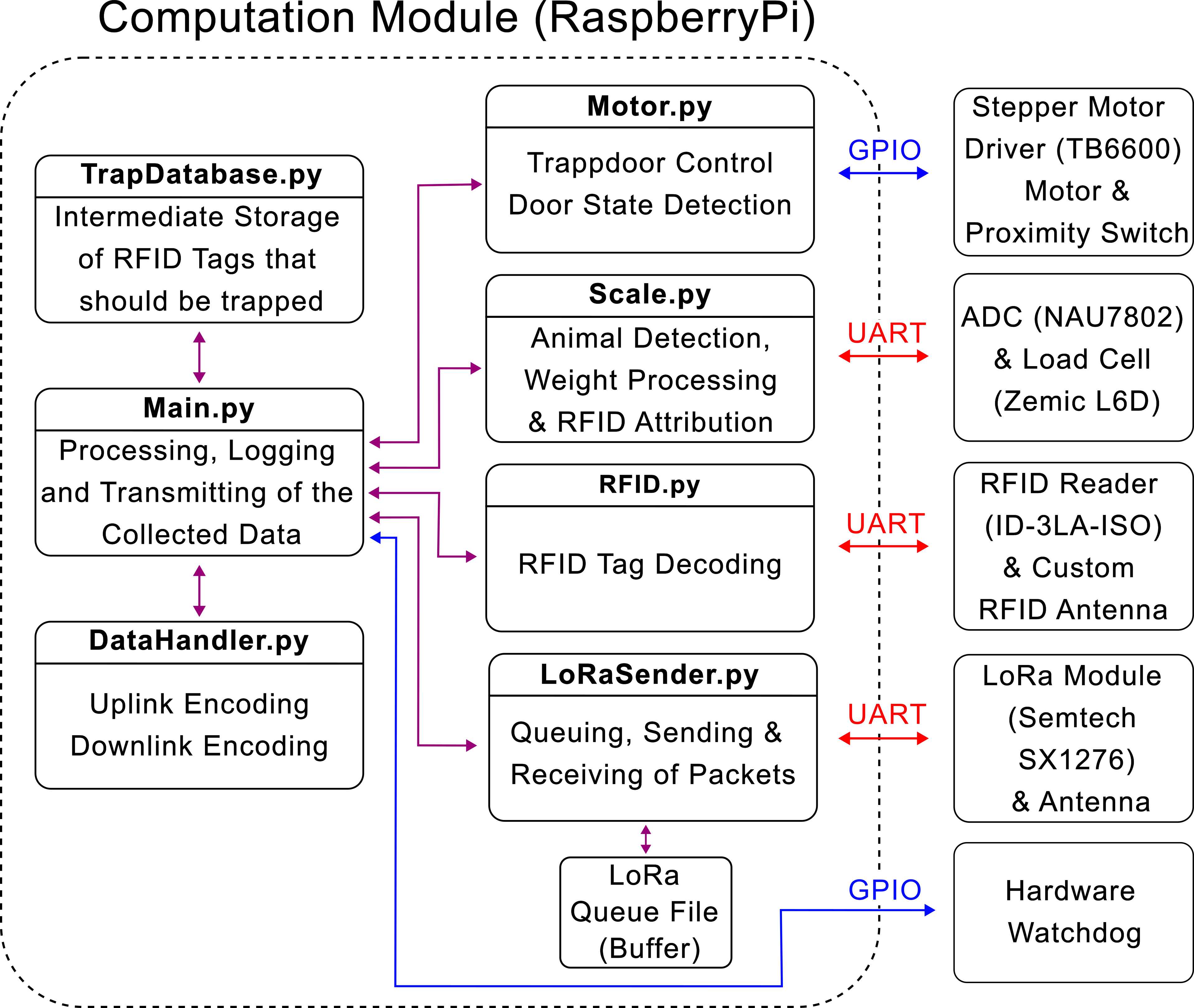}
    \caption{Software overview illustrating the interconnection between the modules, differentiating between hardware (blue, red) and software (purple) connections.}
    \label{fig:software_overview}
\end{figure}

The software architecture is designed in a modular fashion, with the main file as a centerpiece. The individual sub-systems are separated into threads, which can operate independently from each other, handling tasks simultaneously. The main file manages all systems, logging and processing data, and forwarding relevant information between sub-systems. This interconnection is illustrated in Figure \ref{fig:software_overview}. 

\subsubsection{RFID Subsystem}
The \ac{RFID} sub-system continuously monitors the sensor for incoming tags. If a tag is detected, it is decoded and aligned with a timestamp. This tuple will then be passed over to the main thread for processing.

\subsubsection{Trap Database}
A local trap database has been implemented to reduce the amount of LoRa transmissions. This database contains all the \ac{RFID} tags that should trigger the trap upon detection. If unchipped animals should be trapped, the database also includes a master entry triggering the trap accordingly. Lastly, the database stores the time at which it was last updated. This is used by the webserver to determine which \ac{RFID} tags need to be updated.

\subsubsection{Stepper Motor}
The motor sub-system implements the controlling mechanisms for the doors. It controls the stepper motor driver to open and close the doors and is responsible for the calibration based on the proximity switch feedback loop. 

\subsubsection{Data Handler and LoRa Sub-System}
To facilitate a future serialization of this system, a high LoRa transmission efficiency is paramount. This is achieved by using predefined transmission types that implement bitwise encoding. The data handler realizes the encoding and decoding of uplinks and downlinks. The system uses the following transmission types:

\begin{itemize}
    \item System Update: Uplink sent every 10 minutes, containing status information (inside and outside temperature \& humidity as well as error codes).
    \item Animal Update: Uplink sent when the animal leaves, containing all collected measurement data of one animal (multiple updates are sent if more than one animal is entered).
    \item Database Synchronization: Sending an uplink providing the most recent update time and requesting the tags that need to be updated in the local trap database.
    \item Trap Update: Downlink response to entry synchronization request. This contains the current time and any tags that need to be updated in the trap database.
\end{itemize}

After the encoding, the resulting bitstream is sent to the LoRa sub-system, where all packets that need to be sent are stored in a persistent queue. To ensure high transmission reliability, it uses confirmed uplinks, retransmitting a packet until it receives a confirmation from the LoRa network server.

\subsubsection{Weighing Subsystem}
The weighing sub-system has to account for the animal's movement and the common occurrence of multiple animals entering the feeding station. For this, the weighing system features a state-based implementation:
The default state is the idle state, where zeroing the scale continuously correct deviations from the zero point. The system detects an animal entrance and transfers to the entrance state, given a weight shift of more than \qty{20}{\gram} lasting longer than one second. The entrance state increases an animal counter that keeps track of the number of animals inside the system and timestamps the entrance of an animal.
Afterward, the system enters the weighing state, where samples are collected and stored as part of a measurement period.
The measurement period ends on a state change, which occurs if a weight change of more than \qty{20}{\gram} within a moving window is detected. In case of a positive weight change, the system will transition back to the entrance state; otherwise, it will transition back to the exit state. 
Before transitioning, a stable weight is determined by searching the last measurement period for stability windows. A stability window must fulfill the following conditions:

\begin{itemize}
\item Each sample must not deviate by more than one gram from the previous sample. 
\item The previous condition must hold for at least one second of measurements.
\end{itemize}

\begin{figure}[h]
    \centering
    \includegraphics[width=0.85\linewidth]{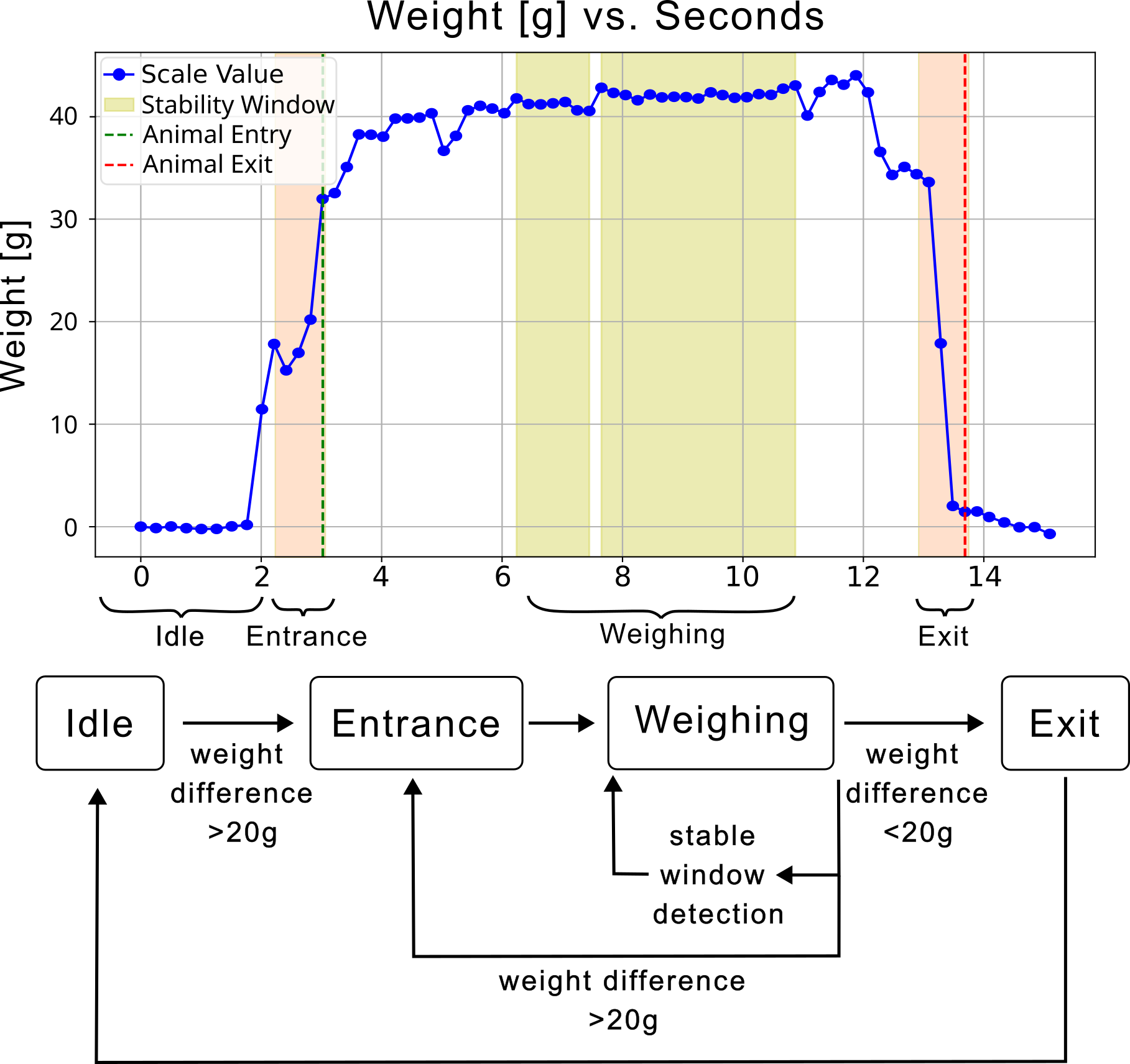}
    \caption{State machine of the weighing system, illustrated with an example measurement. The animal is detected when entering, causing the state machine to transfer from idle to entrance and finally to the weighing state. The weighing state searches for stability windows (green) until an exit state (weight difference larger than \qty{20}{\gram}) is detected.}
    \label{fig:state_machine}
\end{figure}

Finally, only measurements that are part of stability windows will be used to determine the stable weight. The exit state reduces the animal counter and timestamps the exit of an animal. It moves to the idle state if the animal counter reaches zero and otherwise to the weighing state, as there is still an animal present. 
Once the number of animals reaches zero, the animal's weights are determined based on the weight shifts between measurement periods. Finally, the weights are attributed to \ac{RFID} tags that were detected close to the entrance and exit timestamps. 
The collected information is then transmitted through the main file to the LoRa data transmission sub-system and sent via LoRa to the webserver. Figure \ref{fig:state_machine} provides an overview of this state-based weighing approach.

\subsection{Webinterface}
In order to facilitate data extraction and analysis for researchers, the data is stored in a MySQL database, accessible through a website allowing preprocessing (e.g. filtering) and selective data download based on the filtering settings. This minimizes the workload of obtaining the measurement data and enables remote data access. 

\section{ Experimental Evaluation}
On one hand, the experimental evaluation of the designed system consists of a continuous deployment in the Masoala semi-natural rainforest at Zoo Zurich for over two months. On the other hand, multiple laboratory tests were conducted to prove the reliable functionality of individual sub-systems. 

\subsection{Weighing System Evaluation}
To test the accuracy and functionality of the weighing system, a weight test was conducted using three different reference weights of \qty{40}{\gram}, \qty{50}{\gram} and \qty{100}{\gram}. The first two weights were chosen as they represent a mouse lemur under normal circumstances, whilst the last weight represents two animals being on the scale at the same time ~\cite{zoo_mouse_lemur_2024}. With these weights, two tests were conducted: One with a duration of 10 seconds, where the weight was kept stable, and one with a duration of 20 seconds, where the weight was moved to simulate animal movements. In a first step, both tests were performed under laboratory conditions, then repeated after three weeks of deployment to verify the accuracy after continuous use and the impact of a semi-natural rainforest. Table \ref{result_weight_test} provides an overview of all test results. The overall stable weight error of the laboratory tests is \qty{0.45}{\gram} compared to \qty{0.31}{\gram} for the tests when deployed. Nevertheless, the analysis with movable weights revealed an average error of \qty{0.42}{\gram}, verifying the functionality of the animal movement compensation. Overall tests, the average error was determined to be \qty{0.41}{\gram}.

\begin{table}
\centering
\caption{Recorded weighing errors for laboratory and field tests, showing the error when the weight is kept still on the scale compared to when it is moved. This is done for different values representing the lemur's weight.}
\label{result_weight_test}
\begin{tabular}{@{}lrr@{}} 
 \toprule
 \textbf{Measurement Condition} & \textbf{Error Lab-Test} & \textbf{Error Field-Test}\\ 
 \midrule
 \qty{40}{\gram} stable weight for \qty{10}{\second}   & \qty{0.44}{\gram} & \qty{0.36}{\gram} \\
 \midrule
 \qty{40}{\gram} moving weight for \qty{20}{\second}   & \qty{0.41}{\gram} & \qty{0.05}{\gram} \\
 \midrule
 \qty{50}{\gram} stable weight for \qty{10}{\second}   & \qty{0.19}{\gram} & \qty{0.25}{\gram} \\
 \midrule
 \qty{50}{\gram} moving weight for \qty{20}{\second}   & \qty{0.43}{\gram} & \qty{0.23}{\gram} \\
 \midrule
 \qty{100}{\gram} stable weight for \qty{10}{\second}   & \qty{0.94}{\gram} & \qty{0.45}{\gram} \\
 \midrule
 \qty{100}{\gram} moving weight for \qty{20}{\second}   & \qty{0.7}{\gram} & \qty{0.53}{\gram} \\

\bottomrule
\end{tabular}
\end{table}

\subsection{RFID Testing}
To ensure the detection of individual animals, the reliability of the \ac{RFID} scanning must be verified. This is done by moving an \ac{RFID} tag in and out of the entry tube within two seconds. This test was first conducted once under laboratory conditions and later in the field to evaluate the impact of increased humidity on the \ac{RFID} antenna. Tested for 200 samples moving thorough the \ac{RFID} sensor, it revealed to achieve a detection rate of 96.5\% under laboratory conditions and 88.5\% in the field. This results in an overall probability of detecting an animal when visiting the box in the field (entering and leaving) of 98.68\%.

\subsection{Field Testing}
\begin{figure}[h]
    \centering
    \includegraphics[width=1\linewidth]{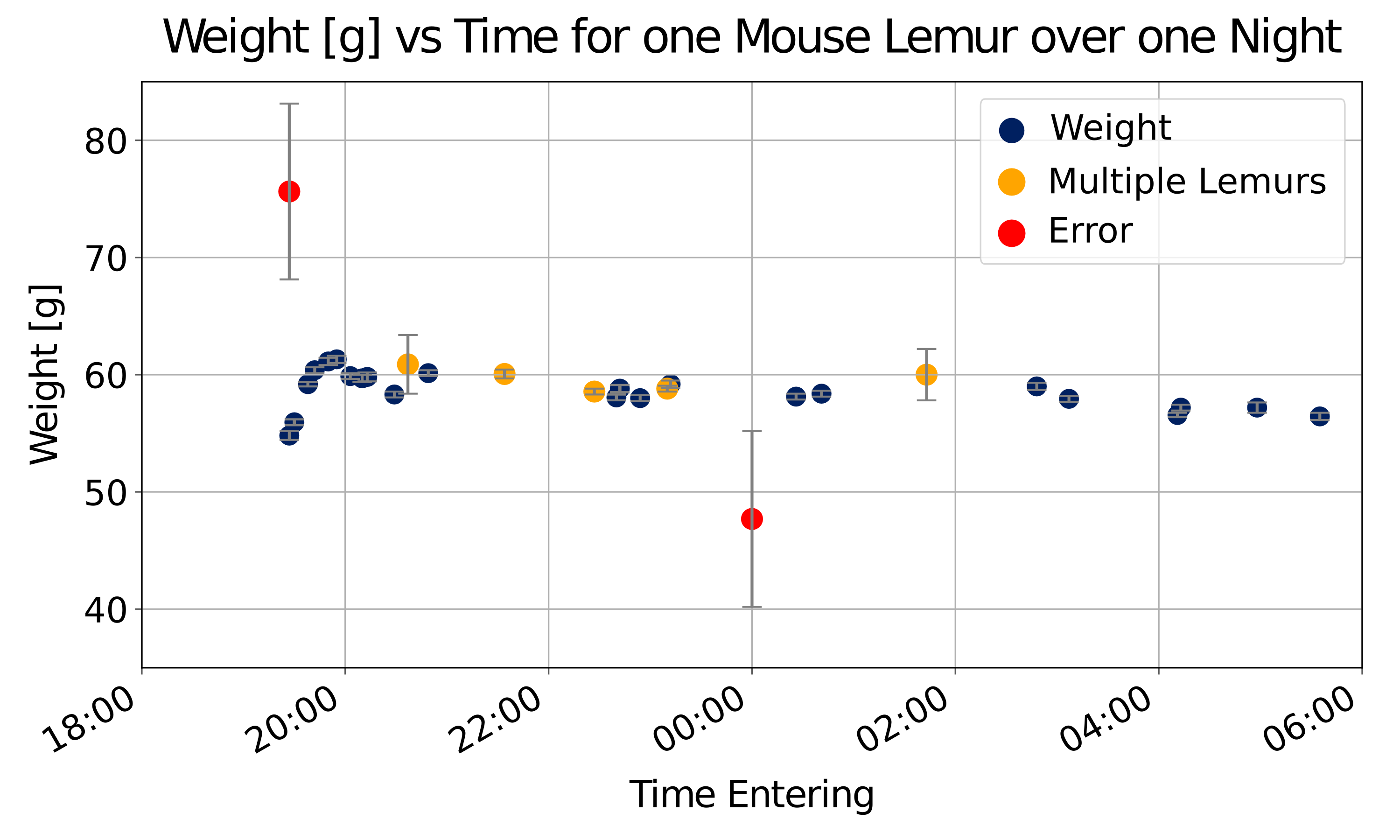}
    \caption{Weight data graph over one night, illustrating how the weight increases at the beginning of the night and indicating instances of two lemurs being present simultaneously.}
    \label{fig:weight_data}
\end{figure}

In a final deployment, the feeding station was tested in the tropical environment of the semi-natural Masoala rainforest at the Zürich Zoo for over two months. The main goal of this evaluation was to test the box against all external influences, such as heavy rain, humidity, and other animals. 
After all, this test should prove that the station will be working as a permanent setup in its final application environment. In total, the box was deployed for 60 days, and during that time, we did not encounter any issues due to external impact. During this testing phase, the LoRa communication reliability was tested as well, proving to reach a reliability of 97.99\% on a total of 1995 transmitted data packets. Additionally, this deployment also allowed us to determine whether the system can be embedded into the workflow of the zookeepers concerning data extraction and maintenance efforts such as cleaning. During the testing, an extensive amount of animal data was collected, providing insights into the quality and potential of the collected data. As an example, Figure \ref{fig:weight_data} plots the weight measurements and visiting times of a specific animal. This data illustrates for one night the frequency of the animal visiting the station as well as its measured weight. Our weighing system attributes the standard deviation of the measurement sequence and marks the quality of the measurement. This allows for post-processing and filtering of data. Nevertheless, such data can be compared between animals, suggesting potential social interaction and group behavior. 

\begin{figure}[h]
    \centering
    \includegraphics[width=1\linewidth]{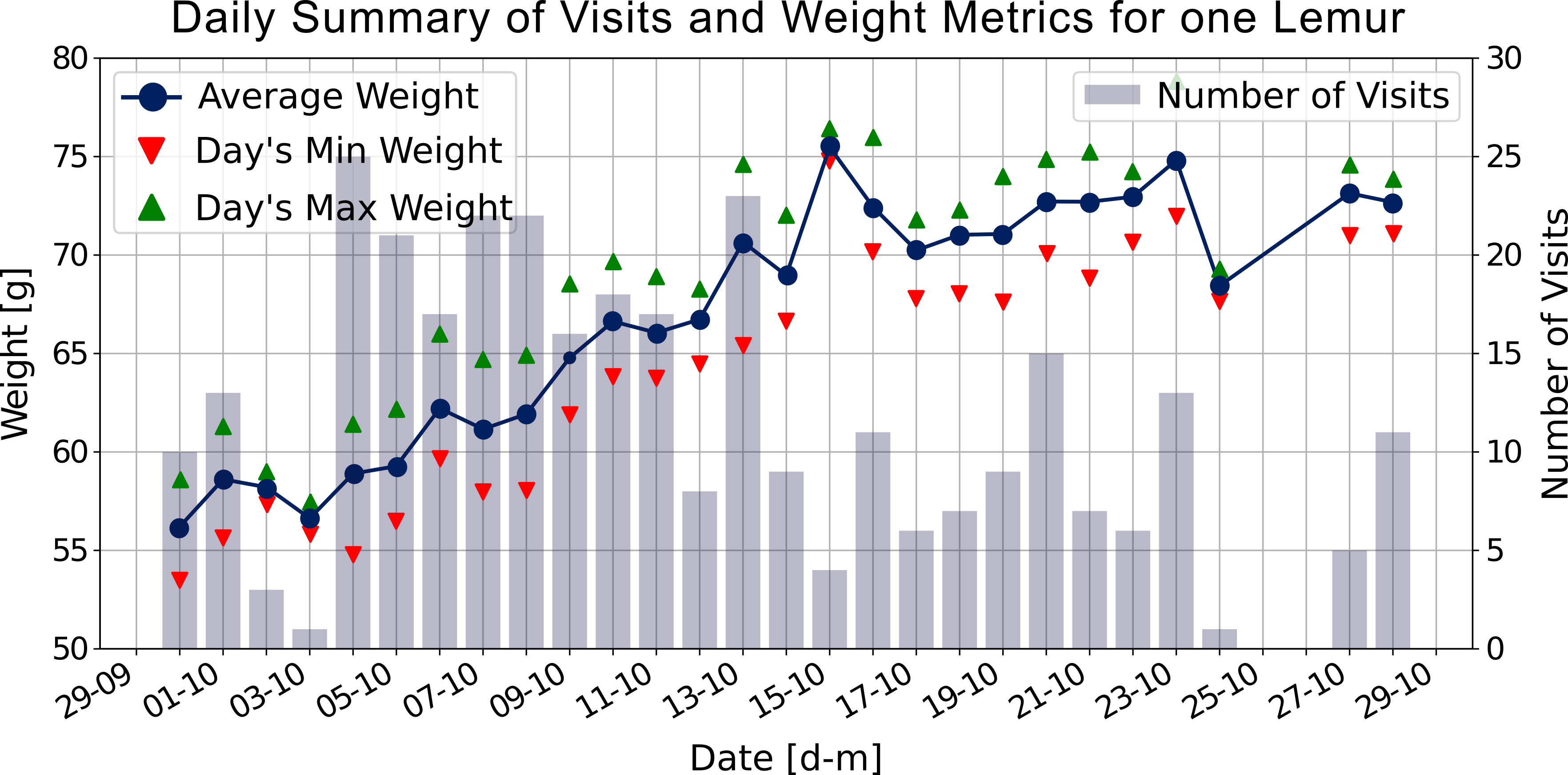}
    \caption{Daily minimum, average, and maximum weights and number of visits for a specific lemur. Showing the weight gain in preparation for winter torpor.}
    \label{fig:daily_weight}
\end{figure}

Finally, the data can be evaluated on a larger time scale, as illustrated in Figure \ref{fig:daily_weight}. The weight measurements and number of visits are described for a specific animal over a period of 29 days. This data shows a spectacular phenomenon, where the animal prepares itself for winter torpor. With this system, such and similar data could be collected for each lemur throughout a lifetime, giving us profound insights into animal health and behavioral patterns.

\section{Conclusion}
This paper presents an automated feeding station to autonomously monitor Goodman's mouse lemurs in a semi-natural rainforest. By incorporating automated animal detection by \ac{RFID} tag scanning along with a weight measurement system to accurately determine the weight of the moving animals, the system allows the collection of essential data necessary for health monitoring and research. Focusing on robustness, minimal maintenance, and streamlined data acquisition through a web interface, the proposed system allows integration into a zoological environment without impacting current work routines. The feeding station has been successfully tested in a semi-natural tropical biome for over two months, exposing it to harsh environmental conditions with periodic rain and high humidity. Over 1000 visits were detected during the field test attributed to 20 different \ac{RFID} tags and multiple animals without a tag. The successful deployment demonstrated how such automated systems can be applied in zoos, enabling them to collect more coherent data about animals, ultimately improving animal welfare. 

\section{Acknowledgments}
We thank the expertise of Zoo Zurich's Masoala team that made this joint prototype possible: Basil von Ah, Francesco Biondi, Daniela Kollmuss, and Angus Sunner. 
We also thank EWZ for providing the \ac{LoRaWAN} infrastructure and Marcus Cathomen for initiating the project and his support.

\begin{acronym}

    \acro{BAN}{Body Area Network}
    \acro{WBAN}{Wireless Body Area Network}
    \acro{PAN}{Personal Area Network}
    \acro{IoT}{Internet of Things}
    \acro{IoB}{Internet of Bodies}
    \acro{AI}{Artificial Intelligence}
    \acro{MCU}{Microcontroller}
    \acro{GNSS}{Global Navigation Satellite System}
    \acro{Nb-IoT}{Narrowband IoT}
    \acro{LoRa}{Long Range}
    \acro{LoRaWAN}{Long Range Wide Area Network}
    \acro{BLE}{Bluetooth Low Energy}
    \acro{UWB}{Ultra-Wideband}
    \acro{NFC}{near Field Communication}
    \acro{SpO2}{Oxigen Saturation}
    \acro{VR}{Virtual Reality}
    \acro{AR}{Augmented Reality}
    \acro{EQS-HBC}{Electro-Quasistatic Human Body Communication}
    \acro{EQS}{Electro-Quasistatic}
    \acro{HBC}{Human Body Communication}
    \acro{WBAN}{Wirelss Body Area Network}
    \acro{WPAN}{Wireless Personal Area Network}
    \acro{WP}{Work Package}
    \acro{CSMA}{Carrier Sense Multiple Access}
    \acro{KB}{Kilobyte}
    \acro{PMIC}{Power Management IC}

    \acro{HCI}{Human-Computer Interaction}
    \acro{HMI}{Human-Machine Interaction}
    \acro{TIA}{Transimpedance Amplifier}
    \acro{SoTA}{State-of-The-Art}
    \acro{WLAN}{Wireless Local Area Network}
    \acro{PCE}{power Conversion Efficiency}
    \acro{ECG}{Electrocardiogram}
    \acro{BOM}{Bill of Material}
    \acro{DAQ}{Data Acquisition}
    \acro{RSSI}{Received Signal Strength Indicator}
    \acro{RSS}{Received Signal Strength}
    \acro{GBP}{Gaind-Bandwidth Product}

    \acro{LoS}{Line-of-Sight}
    \acro{nLoS}{non-Line-of-Sight}
    \acro{QoS}{Quality of Service}
    \acro{NB}{Narrowband Communication}
    \acro{WPT}{Wireless Power Transfer}
    \acro{IBPT}{Intra-Body Power Transfer}
    \acro{ICNIRP}{International Commission on Non-Ionizing Radiation Protection} 
    \acro{RFID}{Radio Frequency Identification}

    \acro{RF}{Radio Frequency}
    \acro{IoT}{Internet of Things}
    \acro{IoUT}{Internet of Underwater Things}
    \acro{UWN}{Underwater Wireless Network}
    \acro{UWSN}{Underwater Wireless Sensor Node}
    \acro{AUV}{Autonomous Underwater Vehicles}
    \acro{UAC}{Underwater Acoustic Channel}
    \acro{FSK}{Frequency Shift Keying}
    \acro{OOK}{On-Off Keying}
    \acro{ASK}{Amplitude Shift Keying}
    \acro{UUID}{Universal Unique Identifier}
    \acro{PZT}{Lead Zirconium Titanate}
    \acro{AC}{Alternating Current}
    \acro{NVC}{Negative Voltage Converter}
    \acro{NVCR}{Negative Voltage Converter Rectifier}
    \acro{FWR}{Full-Wave Rectifier}
    \acro{GPIO}{General Purpose Input/Output}
    \acro{PCB}{Printed Circuit Board}
    \acro{AUV}{Autonomous Underwater Vehicle}
    \acro{IMU}{Intertial Measurement Unit}
    \acro{BLE}{Bluetooth Low Energy}
    \acro{FSR}{Force Sensing Resistor}
    \acro{SiP}{System in Package}
    \acro{SoC}{System on Chip}
    \acro{SpO2}{Oxigen Saturation}
    \acro{PULP}{Parallel Ultra-Low Power}
    \acro{ML}{Machine Learning}
    \acro{ADC}{Analog to Digital Converter}
    \acro{TCDM}{Tightly Coupled Data Memory}
    \acro{GNSS}{Global Navigation Satellite System}
    \acro{IC}{Integrated Circuit}
    \acro{POM}{Polyoxymethylene}
    \acro{RTOS}{Real-Time Operating System}
    \acro{LoRaWAN}{Long Range Wide Area Network}
    \acro{ML}{Machine Learning}
    \acro{IIR}{Infinite Impulse Response}
\end{acronym}

\bibliographystyle{IEEEtran}
\bibliography{sample.bib}

\end{document}